\documentclass[aps,nofootinbib,longbibliography,prd,twocolumn]{revtex4-1}
\usepackage{amsmath,amssymb}
\usepackage[colorlinks,citecolor=blue,linkcolor=blue,urlcolor=blue]{hyperref}
\usepackage{mathrsfs}

\begin{document}

\title{Pre-big-bang black-hole remnants and past low entropy}

\author{Carlo Rovelli${}^1$  and Francesca Vidotto${}^{1,2}$\\[-2mm]{\ }}

\affiliation{$^{1}$  CPT, Aix-Marseille Universit\'e, Universit\'e de Toulon, CNRS, Marseille, France;\\
$^{2}$ University\,of\,the\,Basque\,Country UPV/EHU, Departamento\,de\,F\'isica\,Te\'orica, 
E-48940 Leioa, Spain.}

\date{\small\today}

\begin{abstract}
\noindent Dark matter could be composed by black-hole remnants formed before the big-bang era in a bouncing cosmology. This hypothetical scenario has  implications on the issue of the arrow of time: it upsets a common attribution of past low entropy to the state of the geometry, and suggests a possible realisation of the perspectival interpretation of past low entropy.
\end{abstract}

\maketitle

\section{Remnants}

One of the candidates as dark matter constituent is provided by remnants of evaporated black holes.  This is an old idea \cite{J.H.MacGibbon1987, Barrow1992, Carr:1994ar, Liddle1997, Alexeyev2002,Chen2003, Barrau:2003xp, Chen2004, Nozari2008}, which has  received renewed attention \cite{Rovellia} because of  advances related to quantum gravity \cite{Rovelli2014h,Haggard2014,Christodoulou2016,DeLorenzo2016, Bianchi2018, DAmbrosio2018, Rovelli2018a}. There are no current observational constraints on this possible contribution to dark matter \cite{Carr2016}.

Remnants could be produced at the end of the evaporation of primordial black holes formed at reheating; this is the scenario studied in \cite{Rovellia}.   But there is a second possibility: they could have formed before the big bang, in a cosmological bounce scenario \cite{Brandenberger2016,Agullo2016}. Black hole remnants passing trough the big bounce could have formed in a contracting phase before the current expanding one \cite{Quintin2016,Carr2017}.
 The possibility that dark matter could be formed by matter that has crossed the big bounce has been previously considered. Roger Penrose has coined the name \emph{erebons}, from the Greek god of darkness Erebos, to refer to dark matter particles of Planckian mass crossing over from one eon to the successive one across `dark epochs' \cite{Penrose2017} in cyclic cosmologies \cite{Penrose2012}.  
Here we consider stable \emph{white-hole} remnants of the kind studied in \cite{Rovellia} coming from large pre-bounce black holes,  and we observe that this scenario has interesting implications for thermodynamics.

A debated question in thermodynamics is why entropy was low in the past. Low past entropy is the source of all irreversibility that we see around us, and current established physics takes it as a fact without any consensual explanation.   A hypothesis on the physical reason for past low entropy was put forward in \cite{Rovelli:2015} and is called the hypothesis of the perspectival origin of past low entropy.  According to this hypothesis, the microstate of the universe is generic, but we happen to  belong to a subsystem of the universe that interact with the rest via a reduced set of variables defining a coarse graining with respect to which the (generic) state of the universe had low entropy in the past.  In other words, it is not the microstate of the universe to be `special', it is our perspective on it which is special.  Like the apparent rotation of the cosmos around us, entropy growth is to be understood in terms of our collocation in the universe and our perspective upon in, not in terms of the universe alone. \\[1EM]

In \cite{Rovelli:2015}, this hypothesis lacked a concrete realisation clarifying what could be peculiar in the part of the universe to which we belong. As we explain below, if dark matter is formed by pre-bounce remnants, there is a possible answer to this question. 

In Section 2 we recall the white hole remnant hypothesis and collect the relevant quantitative facts.  In Section 3 we recall the idea of the perspectival source of past low entropy.  In Section 4 we show how  remnant could play a role in making this hypothesis concrete.  

\section{Erebons}

The hypothesis we consider here is that dark matter has a substantial component formed by white-hole remnants generated by the evaporation of black holes before the big bang era in a bounce cosmology.  As shown in \cite{Bianchi2018}, the lifetime of such remnants is of the order
\begin{equation}
\tau_{WH}\sim m_o^4
\end{equation}
in Planck units $\hbar=G=c=1$, where $m_o$ is the mass that originated the black hole that has then given rise to the remnant. (Quantum gravity is locally Lorentz invariant \cite{Rovelli2003,Rovelli:2002vp} and has no preferred time \cite{Rovelli2011h} but the cosmological context determines a preferred frame and a preferred cosmological time variable.)  The internal volume of this remnant when it is formed is also \cite{Bianchi2018},
\begin{equation}
V_{WH}\sim m_o^4. 
\end{equation}
For a remnant to have survived until today, its lifetime must satisfy 
\begin{equation}
\tau_{WH} > T_H
\end{equation}
where $T_H$ is the Hubble time. Therefore its internal volume at the bounce must have been at least 
\begin{equation}\label{volume}
V_{WH}\sim m_o^4 \sim  \tau_{WH} > T_H.   
\end{equation}
If all dark matter is formed by white-hole remnants, the energy density of remnants must be roughly of the order of the matter density $\rho_M$, and this in turn is related to $T_H$ by the Friedmann equation
\begin{equation}
\frac1{T^2_H}\sim \left(\frac{\dot a}{a}\right)^2\sim \rho_M,
\end{equation}
where we neglect factors of order unit. Therefore the current density $\rho$ of remnants is of the order
\begin{equation}
\rho\sim\rho_M\sim \frac1{T^2_H}.
\end{equation}
If we take the mass of each remnant to be Planckian \cite{Bianchi2018}, namely of order unit in Planck units, $\rho$ is also their number density, in the units we are using. 

From the primordial universe to the current epoch, the universe has expanded widely. We can consider approximatively 60 e-foldings for the expansion during the cosmological epochs characterised respectively by radiation, matter, and cosmological constant domination.  In today's most accepted cosmological scenario, one adds a further inflationary expansion of 60 e-folding or more.  Current dark matter is unlikely to be formed by stable remnants that have survived this entire expansion, because their initial density at the bounce would have to be far above the Planck density, and this scenario in tension with the expectation that quantum gravity bounds the density to Planckian values. 

There is however a well-known alternative to inflation in bouncing models: the \emph{matter bounce} scenario \cite{Wands1999,Finelli2002,Allen2004}, where the contracting phase of the bounce is matter-dominated and yields an almost scale-invariant power spectrum  (see \cite{Brandenberger:2012zb,Cai:2014bea} for a review). 
There are different realizations of the matter bounce. 
In particular, the matter bounce has been considered within the bouncing Loop Quantum Cosmology \cite{Wilson-Ewing2013}, where the bounce is produced by loop quantum gravity effects rather than by exotic matter. 
In the literature the matter component is commonly introduced as a scalar field. We note that the presence of black holes is rather generic during a cosmological contraction \cite{Banks:2002fe,Lifshitz:1962bh}, constituting a dust-like preassureless matter content precisely dominating the contraction phase. We consider here a Loop Quantum Cosmology bounce with a black-hole dominated contraction. The details of such model deserve further analysis, our interest here is just to give a preliminary estimation of the viability of this model. 

In a matter bounce scenario, there is no standard inflationary phase.  The quantum bounce yields a super-inflationary phase whose length depends on the model \cite{Mielczarek:2010bh,Ranken:2012hp}. At the bounce the universe is radiation-dominated. The radiation-dominated epoch can be taken to start very close to the bounce \cite{Brandenberger:2012zb}: here for simplicity we consider this case, so that at the bounce the matter and the radiation components have a matter density of the same order of magnitude. In the model studied in \cite{Wilson-Ewing2013}, the density at the bounce is constrained to be of the order 
\begin{equation}
        \rho_b\sim 10^{-9}
\end{equation}
in Planck units.   Using \eqref{volume} we have an amount of internal volume per unit of external volume
\begin{equation}
V_{int}=\rho_b V_{WH}>10^{-9} T_H \sim 10^{52} 
\end{equation}
This means that only a tiny fraction 
\begin{equation}
\frac{1}{V_{int}} < 10^{-52}
\end{equation}
of the volume of the universe was outside the remnants at the bounce!  If we assume equiprobability for each equal volume of the universe, the probability for an observer to be outside those remnants at the bounce is one part in $10^{52}$.   Therefore an observer outside the remnants is in this sense "special" as one part in $10^{52}$. 

The scenario considered here, where black holes cross the bounce, is related but distinct from string cosmology scenarios, such as the one proposed for instance in \cite{Quintin2018:pbh}. There black holes are formed in the high curvature phase, preceding the transition, and they decay very rapidly, while the remnants we are considering are already present in the contracting phase and they are stable for a very long time. 

\section{Past low entropy}

The mystery of the second principle of thermodynamics is not why entropy growth towards the future.  That's pretty obvious.   The mystery is why entropy diminishes going towards the past \cite{Lebowitz1993,Lebowitz,Lanford1981,Uffink2010,Albert2000,Price}. All current irreversible phenomena, including our own future-oriented thinking, the existence of memories and the direction of causality we use to make sense of the world, can be traced to the fact that entropy was low in the past \cite{Reichenbach1958,Rovelli2018}. Since matter was apparently near thermal equilibrium in the past, the low entropy was concentrated on the geometry.  In fact, in standard cosmology geometry is assumed to be nearly homogenous. For the gravitational field, homogeneity is a very low entropy configuration, because gravity tends to clump and generically the evolution of perturbations leads increasingly away from homogeneity. 
In fact, as long argued by Penrose, generic states of gravity, to which generic evolution tends, are highly crumpled, not homogeneous.  (For interesting criticisms and informed alternative perspectives, see \cite{Earman2006,Callender,Callender2004,Wallace2010}.)

The fact that source of past low entropy, hence the source of irreversibility, was the homogeneity of space is confirmed by a simple analysis of the thermodynamical history of the universe.   For instance irreversibility on Earth is due to the strong source of negative energy formed by the sun; the sun in turn was irreversibly formed by the collapse of a primordial cloud under gravitational attraction.   Therefore the original negative entropy driving irreversibility around us can be traced to the early lack of gravitational clumping.  As repeatedly pointed out by Penrose, the fact that the geometry of the universe was small and homogenous to the degree required by the current standard cosmological model, implies a very  `special' state determining an initial low entropy. 

But if the remnant scenario is correct, the geometry at the bounce had far more volume and was not homogenous at all.  To the opposite, it was very highly crumpled. If so, what is the origin of past low entropy, if it is not how special the initial geometry was?

\section{Perspectival entropy}

An imposing aspects of the Cosmos is the mighty daily rotation of Sun, Moon, planets, stars and all galaxies around us.  Why does the Cosmos rotate so?  Well, it is not the Cosmos rotating, it is us. The rotation of the sky is a \emph{perspectival} phenomenon: we understand it better as due to the peculiarity of our own moving point of view, rather than as a global feature of all celestial objects.   The list of conspicuous phenomena that have turned out to be perspectival is long; recognising them has been a persistent aspect of the progress of science. 

The hypothesis put forward in \cite{Rovelli:2015} is that the increase of entropy is a perspectival phenomenon in this sense.  To be sure, it is not subjective or mental, or illusory.  Rather, its source is in the relation between an observer system and an observed system, like for the rotation of the sky. 

This is possible because the entropy of a system depends on the system's microstate \emph{but also} on the coarse graining under which the system interacts. The relevant coarse graining is determined by the concrete existing interactions with the system. The entropy we assign to systems around us depends on the way \emph{we} interact with them -- as the apparent motion of the sky depends on our own motion.  

This observation opens a novel way for facing the puzzle of the arrow of time:  the universe is in a generic state, but sufficiently rich to include subsystems whose coupling defines a coarse graining for which entropy increases monotonically.  These subsystems are those where information can pile up and  `information gathering creatures' such as those composing the biosphere can exist.

A subsystem of the universe that happens to couple to the rest of the universe via macroscopic variables determining an entropy that happens to be low in the past, is a system to which the universe  appears strongly time oriented.  As it appears to us.  Past entropy  may appear low because of our own perspective on the universe.  We refer to \cite{Rovelli:2015} for a detailed discussion of this hypothesis.  

What was not clear in  \cite{Rovelli:2015}, however, is in which sense we  belong to a subsystem of the universe sufficiently special to generate the large past low entropy we see. 

\section{Erebons and the arrow of time}

It is clear at his point that the erebon scenario addresses the issue of the arrow of time.  On the one hand, it replaces the very special \emph{homogeneous} initial state of the conventional Robertson-Walker geometry with a far more generic  "very crumpled" geometry: one where the largest share of the volume is trapped into white-hole remnants. In this way the strange special aspect of the initial state is eliminated or attenuated.   

On the other hand, it shows that by being outside the remnants we are in an extremely non generic situation. Thus providing a possible underpinning of the peculiarity of the coarse graining that our special observational point on the universe implies.  

What we gain in the exchange is a simple explanation of the arrow of time in terms of the very weak (and largely uncontroversial) version of an anthropic argument: in a generic universe, critters like us that exist thanks to a strong entropy gradient are born in those very special regions that allow such entropy gradient.   This is the same as answering to the question of why we are not in a generic spot of the universe -- a generic spot in the universe is in empty extragalactic space: we are not in empty extragalactic space because we are the product of a special universe region with appropriate temperature and a local entropy flow.  Similarly, we come from the extremely tiny region outside the remnants because that's the region with respect to which there is the irreversibility that has made us. 

The reason for the entropic peculiarity of the past, thus, should not be sought in the cosmos at large. Time asymmetry, and therefore  "time flow", might be a feature of the peculiar subsystem to which we belong, a feature needed for information-gathering creatures like us to exist, not a feature of the universe at large. 

This scenario is incomplete.  We have not studied how being outside the remnants determines a coarse graining and why this yield past low entropy.  Here we have simply pointed out the erebon remnant scenario provides ingredients for exploring these possibility. 
\\[2em]

\acknowledgments{
We thank Aureli\'en Barrau for pointing to us a mistake estimating the remnant density at the bounce in the first version of this work, and Robert Brandenberger for his useful insights on the matter bounce.
\\
The work of FV at UPV/EHU is supported by the grant IT956-16 of the Basque Government and by the grant FIS2017-85076-P (MINECO/AEI/FEDER, UE).
}

%

\providecommand{\href}[2]{#2}\begingroup\raggedright\endgroup
\end{document}